# Trajectory-state theory of the Klein-Gordon field

Peter Holland[1]

**Abstract**

We develop a trajectory construction of solutions to the massless wave equation in $n+1$ dimensions and hence show that the quantum state of a massive relativistic system in 3+1 dimensions may be represented by a stand-alone four-dimensional congruence comprising a continuum of 3-trajectories coupled to an internal scalar time coordinate. A real Klein-Gordon amplitude is the current density generated by the temporal gradient of the internal time. Complex amplitudes are generated by a two-phase flow. The Lorentz covariance of the trajectory model is established.

## 1 Introduction

Field theory was developed in the 18th century in the context of extending Newton's discrete particle mechanics to continuous media. Two complementary pictures emerged [1]. In the material, or Lagrangian, picture the state of a continuous system is defined by the displacement function of a continuum of interacting points ('particles') whose spacetime trajectories chronicle the temporal evolution of the state. In the spatial, or Eulerian, picture the state is defined in terms of a few key spacetime functions (such as density and velocity in the case of hydrodynamics) and its temporal evolution is recorded at fixed space points. The two pictures are connected by a well-defined mapping but they are not in one-to-one correspondence either mathematically (the material model may admit a trajectory relabelling symmetry) or conceptually (the pictures pertain to different physical questions, e.g., chaotic phenomena may be exhibited in one and not the other [2]). This disparity indeed justifies the contemplation of alternative notions of state.

     Subsequent applications of the field method beyond continuum physics, in particular to the post-aether electromagnetic field and in general relativity and quantum theory, have relied exclusively upon the spatial description. The author has initiated a programme developing alternative trajectory models of the states of general field systems, beyond those treated in continuum physics but drawing upon its successful methods. This has particular significance in the case of quantum theory, for such a conception challenges two entrenched views: that the quantum state is described by the wavefunction, and that it is forbidden to conceive of simultaneously well-defined position and momentum variables in defining the state. Nevertheless, it has been established that the (spatial) wavefunction picture of the state may be replaced by a material picture in which the burden of temporal evolution is borne solely by a congruence of trajectories, which, moreover, are computed independently of the wavefunction (subject to concordance of the initial conditions in the two pictures) [3,4].

     The quantum trajectory theory turns out to be valuable for another reason. The initial programme set up trajectory pictures for a class of field theories that may be expressed in the form of the Schrödinger equation (first-order in external time, quadratic kinetic energy operator) with a suitably chosen configuration space and Riemannian metric [5,6]. This Schrödinger formulation is efficacious because (a) it equates to a closed system of hydrodynamic-like equations in the spatial picture (via the polar decomposition $\psi = R\exp(iS/\hbar)$ of the wavefunction) and (b) the hydrodynamic fields all have interpretations in a corresponding self-contained trajectory theory, following the tenets of continuum physics. The complementary pictures therefore form a coherent whole. A particular feature is

---

[1] Green Templeton College, Oxford University, peter.holland@gtc.ox.ac.uk

the fundamental role played by the spatial continuity equation, which has a simple solution in terms of trajectories (due to Euler [1]). This approach exhibits a generic role for the quantum potential and embraces trajectory-state models for the many-body Schrödinger equation (which thereby acquires a representation in 3-space [7]), spin ½ systems and quantum fields [6], and non-quantum systems such as the classical electromagnetic field [5] (for further developments see [4]). The significance of the trajectories depends on the context; they may support energy transport, for example, or be more abstract conveyors of probability.

Not all field theories conform to this Schrödinger template, or at least it may not be obvious how to recast them suitably. In that case, we may resort to alternative methods to obtain closed sets of hydrodynamic-like equations. In this endeavour one may exploit a notable feature of some field theories: that they not only imply continuity equations obeyed by nonlinear functions of the field variables, but the field equations *themselves* may be expressed in the form of continuity equations, with linear combinations of field variables or their derivatives representing 'conserved densities'. An example of such a theory has been developed in connection with the Schrödinger equation, which, in addition to the 'standard' form mentioned above, admits a closed continuity-equation format associated with a two-phase trajectory picture [8].

Being second order in the evolution parameter, the Klein-Gordon equation $\partial_\alpha^\alpha \psi + (mc/\hbar)^2 \psi = 0$ falls into the category of 'non-standard' theories just mentioned[2]. Using the polar decomposition, the complex field equation implies the two real equations [10]

$$\partial_\alpha j^\alpha = 0, \quad j^\alpha j_\alpha = c^2 R^4 [1 + (\hbar/mc)^2 \partial_\alpha^\alpha R / R], \quad \alpha = 0,1,2,3. \tag{1.1}$$

These involve the 4-current $j^\alpha = (\rho c, \rho v^r), r = 1,2,3$, with density $\rho = -R^2 \dot{S}/mc^2$ and velocity $v^r = -c^2 \partial_r S / \dot{S}$ but the equations (1.1) are evidently not closed in these hydrodynamic-like variables; the function $R$ also appears alone and it has no obvious interpretation in a putative self-contained trajectory model of the quantum state. On the other hand, as we examine below, the Klein-Gordon equation is an instance of the massless wave equation $\partial_\mu^\mu \Omega = 0$ in one higher dimension, which is a continuity equation obeyed by the current $J^\mu = \partial^\mu \Omega$. This observation invites representing a complex amplitude through its real and imaginary parts rather than its polar decomposition. The corresponding self-contained formulation in terms of density and velocity fields remedies the shortcomings of the approach based on (1.1). Our strategy in devising a trajectory construction of the Klein-Gordon amplitude is therefore to carry this out first for the real wave equation in $n+1$ dimensions, generalizing our previous 1+1 model [11]. This study complements our analysis of the wave equation in material coordinates [12].

In showing that the wavefunction conception of the relativistic quantum state may be replaced by a stand-alone trajectory model, we do not seek to attach a substantive 'massive particle' to one of the paths (which presumably should therefore be timelike, a property not possessed by our model) or to treat them as probability vehicles. Rather, the aim is to explore alternative ways of conceiving the process and of solving the field equation. In this connection, our method attributes state evolution to a continuum of three-dimensional trajectories coupled to an internal time function that evolves with external time. *A real Klein-*

---

[2] There are two well-known ways to write the Klein-Gordon equation in first-order Schrödinger form but neither offers a route to a self-contained trajectory theory: (a) using a two-component wavefunction [9] (requires uninterpreted Pauli matrices); (b) using a scalar time parameter [10] (evolution is not in external time).

*Gordon amplitude is the current density generated by the temporal rate of change of the internal time coordinate.* Complex amplitudes are generated by a two-phase flow. We also establish the Lorentz covariance of the Klein-Gordon trajectory model.

## 2 The wave equation in *n*+1 dimensions

We consider a space of *n* dimensions equipped with Cartesian coordinates $x^i$, $i,j,\ldots = 1,\ldots,n$, and metric $g^{ij} = g_{ij} = \zeta_i \delta_{ij}$, $\zeta_i = \pm 1$, so that $|\det g_{ij}| = 1$, which is used to raise and lower indices. In the trajectory picture, the state of a continuous system embedded in the space is described through the functional dependence of the displacement vector $q^i(a,t)$ of a point on the time *t* and on the trajectory's distinguishing coordinates $a^i$ (which will be identified with the position at $t = 0$: $q^i(a,t=0) = a^i$). We assume that the mapping between the two sets of coordinates $q^i$, $a^i$ is single-valued, differentiable with respect to $a^i$ and *t*, and invertible.

Let $\rho_0(a)$ be the initial density of some continuously distributed quantity in the space (this may be of either sign). Then the quantity in an elementary volume $d^n a$ attached to the point $a^i$ is given by $\rho_0(a) d^n a$. The conservation of this quantity in the course of the motion is expressed through the relation

$$\rho(q(a,t),t) = J^{-1}(a,t) \rho_0(a) \tag{2.1}$$

where

$$J = \det(\partial q / \partial a) = \frac{1}{n!} \varepsilon_{i_1 \ldots i_n} \varepsilon^{j_1 \ldots j_n} \frac{\partial q^{i_1}}{\partial a^{j_1}} \ldots \frac{\partial q^{i_n}}{\partial a^{j_n}}, \quad 0 < J < \infty. \tag{2.2}$$

We seek a force law whose spatial image is the massless wave equation. Generalizing our previous 1+1-dimensional treatment [11], we consider the following second-order (in *t* and $a^i$) law of motion in *n*+1 dimensions:

$$\rho_0 \ddot{q}^i = J \frac{\partial}{\partial q^j} \left[ \rho_0 J^{-1} \left( c^2 g^{ij} + \dot{q}^i \dot{q}^j \right) \right] \tag{2.3}$$

where $\dot{q}^i = \partial q^i(a,t)/\partial t$,

$$\frac{\partial}{\partial q^i} = J^{-1} J_i^j \frac{\partial}{\partial a^j}, \tag{2.4}$$

and $J_i^j$ is the adjoint of the deformation matrix $\partial q^i / \partial a^j$ with

$$\frac{\partial q^i}{\partial a^j} J_i^l = J \delta_j^l, \quad J_i^j = \frac{\partial J}{\partial (\partial q^i / \partial a^j)}. \tag{2.5}$$

In this picture $\partial / \partial t$ is calculated for constant $a^i$. The initial conditions to be appended to the dynamical equations (2.3) are assumed to be connected to the Cauchy data (denoted $\Psi_0(x), \partial \Psi_0(x)/\partial t$ ) for the wave whose evolution is to be simulated by the congruence, via the relations

$$\rho_0(a) = c^{-2} \partial \Psi_0(a)/\partial t, \quad \rho_0(a)\dot{q}_{i0}(a) = \partial \Psi_0(a)/\partial a^i. \tag{2.6}$$

An important property of the law (2.3) is that it implies, in conjunction with (2.6), that the velocity current vector is irrotational for all $t$. This is proved below in the spatial language.

Given the initial data, (2.3) is a self-contained field equation in $q^i$. The density- and velocity-dependent force we have postulated is justified by the description it implies in the spatial picture, obtained by mapping the dependent variables $q^i(a,t)$ into independent variables $x^i$. The density and velocity are obtained from the following formulas:

$$\rho(x,t) = J^{-1}(a,t)\rho_0(a)\big|_{a(x,t)}, \quad v^i(x,t) = \dot{q}^i(a,t)\big|_{a(x,t)}. \tag{2.7}$$

Differentiating each relation (2.7) with respect to $t$ and using (2.3) we easily deduce the four field equations

$$\frac{\partial \rho}{\partial t} + \partial_i(\rho v^i) = 0 \tag{2.8}$$

$$\rho\left(\frac{\partial v^i}{\partial t} + v^j \partial_j v^i\right) = \partial_i\left[\rho(c^2 g^{ij} + v^i v^j)\right] \tag{2.9}$$

where $\partial_i = \partial/\partial x^i$ and in this picture $\partial/\partial t$ is calculated for constant $x^i$. The initial conditions implied by (2.6) are

$$\rho_0(x) = c^{-2} \partial \Psi_0(x)/\partial t, \quad \rho_0(x)v_{i0}(x) = \partial_i \Psi_0(x). \tag{2.10}$$

These equations are self-contained in $\rho$ and $v^i$, (2.8) being the continuity equation and (2.9) the spatial version of Euler's law, corresponding to a pressure tensor $p^{ij}(\rho,v) = -\rho(c^2 g^{ij} + v^i v^j)$. Using the continuity equation, the force law can be rewritten as

$$\frac{\partial}{\partial t}(\rho v^i) - c^2 g^{ij} \partial_j \rho = 0. \tag{2.11}$$

Note that there are no constraints on the sign of $\rho$ or the magnitude of $v^i$.

To show that the combination of (2.8) and (2.11) is equivalent to the scalar wave equation, we first integrate (2.11) with respect to $t$ to get $\rho v_i - \rho_0 v_{i0} = \partial_i \int_0^t c^2 \rho dt$. Employing the second of the initial conditions (2.10) shows that locally there exists a function $\Psi(x,t)$ such that $\Psi(x, t=0) = \Psi_0(x)$ and $\rho v_i$ is its gradient field for all $t$:

$$\rho v_i = \partial_i \Psi. \tag{2.12}$$

Next, inserting the latter result in (2.11) and integrating with respect to $x^i$, we get $c^2\rho = f(t) + \partial\Psi/\partial t$. Setting the arbitrary function $f = 0$, the first of the initial conditions (2.10) gives $\partial\Psi(x,t)/\partial t|_{t=0} = \partial\Psi_0(x)/\partial t$ and so, for all $t$,

$$c^2\rho = \partial\Psi/\partial t. \qquad (2.13)$$

We unite these results using a condensed notation. Let us write $x^\mu = (ct, x^i)$, $\mu = 0,1,...,n$, and introduce the current vector $J^\mu$ defined by $J^0 = \rho c$, $J^i = \rho v^i$ with $J_\mu = g_{\mu\nu} J^\nu = (\rho c, \rho g_{ij} v^j)$ and

$$g_{\mu\nu} = g^{\mu\nu} = \begin{pmatrix} 1 & 0 \\ 0 & g_{ij} \end{pmatrix}. \qquad (2.14)$$

Transferring to this notation, we have shown that the Euler law (2.11) subject to (2.10) implies $J_\mu = \partial_\mu \Psi$. Noting that the continuity equation (2.8) may be written $\partial_\mu J^\mu = 0$, we deduce that the current potential $\Psi$ obeys the scalar wave equation in $n+1$ dimensions,

$$g^{\mu\nu} \partial_{\mu\nu} \Psi = 0, \qquad (2.15)$$

with Cauchy data $\Psi_0(x), \partial\Psi_0(x)/\partial t$. This is a consequence of the trajectory law (2.3) plus the initial conditions (2.6), the connection being demonstrated through the mediation of the density and velocity functions (2.7). The congruence may therefore be employed to construct the time-dependence of the real scalar wave amplitude via the formula:

$$\begin{aligned} \Psi(x,t) &= \int J_0 c\, dt + J_i dx^i \\ &= \int (c^2 J^{-1} \rho_0)_{a(x,t)} dt + g_{ij} (\dot{q}^i J^{-1} \rho_0)_{a(x,t)} dx^j. \end{aligned} \qquad (2.16)$$

Conversely, we may start from (2.15), construct the relations (2.8) and (2.9), and hence deduce the trajectory equation (2.3). The latter is subject to the material equvalent of the irrotational condition (2.12), namely,

$$\rho_0 J^{-1} g_{jk} \dot{q}^j \frac{\partial q^k}{\partial a^i} = \frac{\partial \Phi(a,t)}{\partial a^i} \qquad (2.17)$$

where $\Phi(a,t) = \Psi(x(a,t),t)$.

### 3 The Klein-Gordon equation

Our quarry, a derivation of massive 3+1 Klein-Gordon evolution from a trajectory model, is achieved by specializing the preceding treatment of the massless wave equation. To this end, we restrict to a four-dimensional space with $q^i = (q^r, q^4)$, $a^i = (a^r, a^4)$ with $r, s,\ldots = 1,2,3$ and metric $g_{ij} = \text{diag}(-1,-1,-1,1)$. The fact that $g_{44} = +1$ suggests that the variable $q^4/c$ plays the part of an additional 'time' coordinate that is dynamical and evolves with the external time $t$. We seek solutions to the trajectory law (2.3) for which

$$q^r(a^s, a^4, t) = q^r(a^s, t), \quad q^4(a^s, a^4, t) = a^4 + c\tau(a^s, t), \quad r, s = 1, 2, 3, \tag{3.1}$$

where $\tau(t=0) = 0$. In addition, we restrict the initial conditions (2.6) by factoring out the $a^4$ dependence,

$$\Psi_0 = \psi_0(a^r)\exp(mca^4/\hbar), \quad \partial\Psi_0/\partial t = \exp(mca^4/\hbar)\partial\psi_0(a^r)/\partial t, \tag{3.2}$$

so that

$$\left.\begin{array}{l}\rho_0(a^r, a^4) = \exp(mca^4/\hbar)\bar{\rho}_0(a^r), \quad \bar{\rho}_0(a^r) = c^{-2}\partial\psi_0(a^r)/\partial t, \\ \bar{\rho}_0(a^r)\dot{q}_{r0}(a^s) = \partial\psi_0(a^s)/\partial a^r, \quad \bar{\rho}_0(a^r)\dot{\tau}_0(a^r) = (m/\hbar)\psi_0(a^r).\end{array}\right\} \tag{3.3}$$

Here $m$ is a constant that we will show may be identified with mass. The law (2.3) now implies the four coupled equations

$$\left.\begin{array}{l}\rho_0 \ddot{q}^r = J\dfrac{\partial}{\partial q^s}\left[\rho_0 J^{-1}\left(-c^2\delta^{rs} + \dot{q}^r\dot{q}^s\right)\right] + \dfrac{mc^2}{\hbar}\rho_0\dot{q}^r\dot{\tau} \\ \rho_0 \ddot{\tau} = J\dfrac{\partial}{\partial q^s}\left(\rho_0 J^{-1}\dot{\tau}\dot{q}^s\right) + \dfrac{mc^2}{\hbar}\rho_0\left(1 + \dot{\tau}^2\right)\end{array}\right\} \tag{3.4}$$

where

$$\frac{\partial}{\partial q^r} = J^{-1}\left(J_r^s\frac{\partial}{\partial a^s} + J_r^4\frac{\partial}{\partial a^4}\right), \quad \frac{\partial}{\partial q^4} = \frac{\partial}{\partial a^4}, \tag{3.5}$$

and we have used the results $J_4^r = 0$ and $J_4^4 = J = \det(\partial q^r/\partial a^s)$ (the 3-determinant).

We assert that *equations (3.4) subject to (3.3) constitute the material version of the Klein-Gordon equation for a quantum system of mass m in the spatial picture*. In the spatial formulation we write $x^\mu = (ct, x^i)$ with $\mu = 0, 1, 2, 3, 4$ and $x^i = (x^r, x^4), r = 1, 2, 3$. The conditions (3.1) and (3.3) imply that the functions (2.7) have the form

$$\begin{array}{l}\rho(x^r, x^4, t) = \exp(mcx^4/\hbar)\bar{\rho}(x^r, t), \\ \bar{\rho}(x^r, t) = [J(a^s, t)^{-1}\bar{\rho}_0(a^s)\exp(-mc^2\tau(a^s, t)/\hbar)]\big|_{a^s(x^r,t)}\end{array} \tag{3.6}$$

$$v^r(x^r, t) = \dot{q}^r(a^s, t)\big|_{a^s(x^r,t)}, \quad v^4(x^r, t) = c\dot{\tau}(a^s, t)\big|_{a^s(x^r,t)}, \tag{3.7}$$

so that the velocity fields $v^r, v^4$ are independent of $x^4$. Combining (3.6) and (3.7) with (2.12) and (2.13) shows that the effect of conditions (3.1) and (3.3) is to factorize the 4+1 amplitude $\Psi$ into a 3+1 function of $x^r$ times a fixed function of $x^4$ for all $t$, up to an additive constant $k$: $\Psi = \psi(x^r,t)\exp(mcx^4/\hbar) + k$. Putting $t = 0$, (3.2) shows that $k = 0$, $\psi(t=0) = \psi_0(a^r)$ and $\partial\psi/\partial t|_{t=0} = \partial\psi_0(a^r)/\partial t$. We now have for $\Psi$ [3,4]

$$g^{\mu\nu}\partial_{\mu\nu}\Psi = 0, \quad \Psi(x^r,x^4,t) = \psi(x^r,t)\exp(mcx^4/\hbar), \quad \mu,\nu = 0,1,2,3,4. \quad (3.8)$$

As a final step, the relations (3.8) together with (3.2) are equivalent to the Klein-Gordon equation for a system of mass $m$ obeyed by the real function $\psi(x^r,t)$,

$$\frac{1}{c^2}\frac{\partial^2\psi}{\partial t^2} - \partial_{rr}\psi + \frac{m^2c^2}{\hbar^2}\psi = 0, \quad r = 1,2,3, \quad (3.9)$$

with the initial conditions $\psi_0(x^r), \partial\psi_0(x^r)/\partial t$.

Using this factorization, (2.12) and (2.13) become five relations in 3+1 dimensions:

$$\bar{\rho}c = \partial_0\psi, \quad \bar{\rho}v_r = \partial_r\psi, \quad \bar{\rho}v_4 = (mc/\hbar)\psi. \quad (3.10)$$

The material irrotationality condition (2.17) becomes the four relations

$$\bar{\rho}_0 J^{-1}\left(c^2\dot{\tau}\frac{\partial\tau}{\partial a^r} - \dot{q}^s\frac{\partial q^s}{\partial a^r}\right) = \frac{\partial\phi}{\partial a^r}, \quad \bar{\rho}_0 J^{-1}\dot{\tau} = (m/\hbar)\phi \quad (3.11)$$

where we write $\Phi(a,t) = \phi(a^r,t)\exp(mca^4/\hbar)$ with $\phi(a^r,t) = \psi(x^r(a^s,t),t)\exp(mc^2\tau(a^s,t)/\hbar)$.

Combining the last relations in (3.10) and (3.11), we have achieved the following construction of a real time-dependent Klein-Gordon amplitude from the trajectories:

$$\begin{aligned}\psi(x^r,t) &= (\hbar/mc)\bar{\rho}(x^r,t)v^4(x^r,t) \\ &= (\hbar/m)[J(a^s,t)^{-1}\bar{\rho}_0(a^s)\exp(-mc^2\tau(a^s,t)/\hbar)\dot{\tau}(a^s,t)]\Big|_{a^s(x^r,t)}.\end{aligned} \quad (3.12)$$

Conversely, given $\psi$, the paths (3.1) may be constructed from the first-order equations (3.7) where, using (3.10), $v_r = c\partial_r\psi/\partial_0\psi$ and $v^4 = (mc^2/\hbar)\psi/\partial_0\psi$.

It is straightforward to obtain a complex solution $\psi_1 + i\psi_2$ of the free Klein-Gordon equation by this method since the equivalence with the higher-dimension massless wave equation extends to this case. The real functions $\psi_1, \psi_2$ are independent solutions and hence their time dependences may be derived from corresponding independent congruences $q_1^i, q_2^i$. Each point in the 4-space then supports two trajectories. In the case of external electromagnetic potentials, the functions $\psi_1, \psi_2$ become coupled and the trajectory laws for $q_1^i, q_2^i$ are correspondingly coupled. Coupled trajectory equations have been treated previously [8]; since the present case does not present any essentially novel features, we do not go into this here.

---

[3] It has been proved [13] that the current $j^\alpha$ of Sec. 1 is the only conserved Lorentz 4-vector implied by the Klein-Gordon equation that is a local function of just $\psi$ and its first derivatives, up to multiplicative and additive constants. The existence of the conserved 5-vector $\partial_\mu\Psi$ is compatible with this theorem.

[4] The interpretation of the exponential growth of the factorized solution as $x^4 \to \infty$ is an open question.

# 4 Lorentz covariance of the Klein-Gordon trajectory equations

In the space $(x^i,t)$, $i = 1,2,3,4$, we consider an infinitesimal Lorentz transformation (with boost velocity $u^r$, $|u^r| \ll c$) of the subset of variables $(x^r,t)$, with respect to which $\psi$ is a scalar ($r,s = 1,2,3$):

$$x'^r = x^r - u^r t, \quad t' = t - u^r x^r / c^2, \quad x'^4 = x^4, \quad \psi'(x'^r,t') = \psi(x^r,t). \qquad (4.1)$$

These relations imply $\Psi'(x'^r,x'^4,t') = \Psi(x^r,x^4,t)$, $\rho' = \rho - u^r \rho v^r / c^2$, $\rho' v'^r = \rho v^r - u^r \rho$ and $\rho' v'^4 = \rho v^4$ from which it is obvious that equations (3.8)-(3.10) of the spatial formalism are covariant.

The corresponding infinitesimal substitution in the material picture is effected in the space $(a^i,t)$ and on functions therein:

$$\left.\begin{array}{l} a'^r = a^r, \quad a'^4 = a^4, \quad t' = t - u^r q^r(a^s,t)/c^2, \quad q'^r(a'^s,t') = q^r(a^s,t) - u^r t, \\ q'^4(a'^s,a'^4,t') = q^4(a^s,a^4,t), \quad \rho'_0(a'^s,a'^4) = \rho_0(a^r,a^4), \quad \phi'(a'^s,t') = \phi(a^s,t). \end{array}\right\} \qquad (4.2)$$

The Lorentz transformation is evidently now state-dependent, while each trajectory $(a^r,a^4)$ transforms into itself (no 'relativity of label' [11]). These formulas imply the following relations:

$$\left.\begin{array}{l} \dfrac{\partial}{\partial a'^r} = \dfrac{\partial}{\partial a^r} + \dfrac{u^s}{c^2}\dfrac{\partial q^s}{\partial a^r}\dfrac{\partial}{\partial t}, \quad \dfrac{\partial}{\partial t'} = \left(1 + \dfrac{u^r}{c^2}\dot q^r\right)\dfrac{\partial}{\partial t}, \\ \dfrac{\partial}{\partial q'^r} = \dfrac{\partial}{\partial q^r} + \dfrac{u^r}{c^2}\left(\dfrac{\partial}{\partial t} - \dot q^s \dfrac{\partial}{\partial q^s} - \dot q^4 \dfrac{\partial}{\partial q^4}\right), \quad \dfrac{\partial}{\partial q'^4} = \dfrac{\partial}{\partial q^4}. \end{array}\right\} \qquad (4.3)$$

Hence

$$\dot q'^r = \dot q^r - u^r + u^s \dot q^s \dot q^r / c^2, \quad \dot\tau' = \dot\tau(1 + u^r \dot q^r / c^2), \quad J' = J(1 + u^r \dot q^r / c^2). \qquad (4.4)$$

Using these results it is straightforward to prove the covariance of the density relation (2.1) and the equations of motion (3.4). The latter proof requires the identity $\partial \log J / \partial t = \partial \dot q^r / \partial q^r$ and the irrotationality condition (3.11). As noted previously, (3.11) follows from the law of motion and the initial conditions (2.6), and it is itself covariant[5].

The condition $q^r(a,t=0) = a^r$ is not preserved by (4.2) but it may be maintained in all frames if the transformation is accompanied by a $u^r$-dependent transformation of the trajectory labels $a^r$, in total a 'label-dependent Lorentz transformation'[6]. To see this, put $t' = 0$ in (4.2) and solve $t = u^r q^r(a^s,t)/c^2$ to get the corresponding time $t(u^r,a^r)$. Then, from (4.2), $q'^r(a'^s, t' = 0) = q^r(a^s, t(u^r, a^r))$ in this approximation. Denoting this last vector function by $a'^s(u^r, a^r)$ provides the desired label mapping. Likewise, we have $q'^4(a'^s, a'^4, t' = 0) = q^4(a^s, a^4, t(u^r, a^r))$, which defines $a'^4(u^r, a^r, a^4)$. Under this label transformation $\rho_0$ transforms as a scalar density while $\rho_0 d^4 a$ and $\rho_0 J^{-1}$ are invariants.

---

[5] The proof of the Lorentz covariance of (3.4) using (3.11) is analogous to that of the dynamical ('curl') Maxwell equations, which requires the kinematical ('div') equations whose validity is likewise implied by the dynamical equations for all $t$ if they hold at an instant.

[6] This is analogous to making a 'gauge-dependent Lorentz transformation' of the electromagnetic potentials to maintain a non-Lorentz-covariant gauge condition in all inertial frames.

## 5 Conclusion. The Klein-Gordon amplitude as a temporal current

We have established a relativistically covariant trajectory flow in 4+1 dimensions that generates real solutions $\psi(x^r, t)$ of the 3+1 Klein-Gordon equation. The 4-dimensional congruence comprises a continuum of 3-trajectories with displacement function $q^r(a^s, t)$, $r,s = 1,2,3$, in the space of the amplitude's argument $x^r$, coupled with what we have deemed an internal time $\tau(a^r, t)$, a Lorentz scalar associated with, and varying along, each 3-trajectory $a^r$. The internal time characterizes the massive nature of the field ($\tau = 0$ when $m = 0$ and the theory reduces to that of Sec. 2 with $n = 3$).

The congruence shares three key properties with the 'standard' Schrödinger approach described in Sec.1: it depends on the form of the wavefunction and not its absolute magnitude (the velocity components in (3.10) are insensitive to multiplication of $\psi$ by a constant); points are linked by at most a single orbit (and not 'all possible paths'); and it exhibits a relabelling symmetry [14]. The approaches diverge in other respects. The force law (2.3) does not employ the quantum potential, and the key role of the fourth coordinate $\tau$ in state propagation is not anticipated in the standard method. Regarding the latter, it is evident from (3.12) that $\psi$ may be expressed as a 'temporal current density', proportional to the rate of change of $\tau$ with external time. To represent this it is useful to replace $\tau$ by the non-negative Lorentz scalar

$$\mathrm{T}(a^r, t) = \alpha e^{-\tau(a^r, t)/\alpha}, \quad \mathrm{T}_0 = \alpha = \hbar/mc^2, \quad \dot{\mathrm{T}}_0(a^r) = -\psi_0/\alpha\dot{\psi}_0. \qquad (5.1)$$

Then, writing (3.12) in 'propagator' form and substituting for $\bar{\rho}_0(a^r)$ from (3.3), the real Klein-Gordon amplitude may be written

$$\psi(x^r, t) = -\alpha \int \dot{\mathrm{T}}(a^s, t)\delta(x^r - q^r(a^s, t))\dot{\psi}_0(a^s) d^3 a. \qquad (5.2)$$

A complex amplitude is generated by two congruences via a superposition:

$$\psi_1 + i\psi_2 = -\alpha \int \left[\dot{\mathrm{T}}_1 \delta(x^r - q_1^r)\dot{\psi}_{10} + i\dot{\mathrm{T}}_2 \delta(x^r - q_2^r)\dot{\psi}_{20}\right] d^3 a. \qquad (5.3)$$

The time function $\mathrm{T}$ may play the role of an internal 'clock' but it need not directly reflect the periodicity of the wavefunction. Consider a plane wave $\psi(x^r, t) = \cos(\omega t - k^r x^r)$ where $\omega^2/c^2 - k^r k^r = m^2 c^2/\hbar^2$. A simple computation using (3.7) gives for the 3-trajectories $q^r = a^r + k^r c^2 t/\omega$ and hence $\mathrm{T} = \alpha |\sin(t/\alpha^2 \omega - a^r k^r)|/|\sin a^r k^r|$. $\mathrm{T}$ is thus a periodic function of $t$ but the frequency $1/\alpha^2 \omega \neq \omega$.

[1] C. Truesdell and R.A. Toupin, in *Handbuch der Physik*, Band III/I, Ed. S. Flügge (Springer-Verlag, Berlin, 1960)
[2] A. Crisanti, M. Falcioni, A. Vulpiani and G. Paladin, *Riv. Nuovo Cim.* 14, 1 (1991)
[3] P. Holland, *Ann. Phys. (NY)* 315, 503 (2005)

[4] P. Holland in *Quantum Structural Studies*, Eds. R. E. Kastner *et al.* (World Scientific, London, 2017)
[5] P. Holland, *Proc. R. Soc. A* 461, 3659 (2005)
[6] P. Holland, in *Quantum Trajectories*, Ed. P. Chattaraj (Taylor & Francis/CRC, Boca Raton, 2010)
[7] P. Holland, *J. Mol. Model.* 24, 269 (2018)
[8] P. Holland, *J. Phys. A: Math. Theor.* 42, 075307 (2009)
[9] H. Feshbach and F. Villars, *Rev. Mod. Phys.* 30, 24 (1958)
[10] P.R. Holland, *The Quantum Theory of Motion* (Cambridge Uni Press, Cambridge, 1993)
[11] P. Holland, *Int. J. Theor. Phys.* 51, 667 (2012)
[12] P. Holland, *Ann. Phys. (NY)* 351, 935 (2014)
[13] P. Holland, *Ann. Phys. (Leipzig)* 12, 446 (2003)
[14] P. Holland, in *Concepts and Methods in Modern Theoretical Chemistry: Statistical Mechanics*, eds. S.K. Ghosh and P.K. Chattaraj (Taylor & Francis/CRC, Boca Raton, 2013)